
%
%
%
%
%
%
%
%
%
%
\font\Btk=cmr10 scaled\magstep3
\magnification=1200
\parskip=5pt plus 2pt
\vsize=9.1 true in
\hsize=6.2 true in
\parindent=20pt
\baselineskip=17pt


\newcount\notenumber
\def\resetnotenumber{\notenumber=1}
\def\note#1{{\baselineskip=12pt\footnote{$^{\the\notenumber}$}{#1}}
\advance\notenumber by1}
\resetnotenumber

\def\ie{{\it i.e.,\ }}


\def\R{{\rm I\! R}}
\def\\C{{\rm C}}
\def\C{\mkern1mu\raise2.2pt\hbox{$\scriptscriptstyle|$}{\mkern-7mu\rm C}}

\def\and{{\rm\ and\ }}

\def\abs #1{\vert {#1}\vert}
\def\frac#1#2{{#1 \over #2}}
\def\half{{\frac 12}}

\def\pde #1#2{\partial_{#2} #1}


\def\a {\alpha}
\def\b {\beta}
\def\g {\gamma}

\def\om {\omega}

\def\f {\psi}

\def\l {\lambda}

\def\r {\rho}

\def\t {\tau}
\def\vf {\chi}
%
%

\def\x {\xi}
\def\y {\eta}

\def\Tr {\rm Tr}
\def\Pe {\rm P}
\def\R  {\cal R}
\def\So {\cal S}
\def\z {\phantom z}
\def\one {\rm I}
%
%
{\nopagenumbers
\rightline{SU-GP-93/7-8}
\rightline{UM-P-93/77}
\rightline{gr-qc/9308021}
\rightline{August 1993}
\vskip 0.5 true in
\centerline{\Btk Integrals of Motion in the Two Killing}
\centerline{\Btk Vector Reduction of General Relativity}
\bigskip
\centerline{Nenad Manojlovi\'c}
\centerline{\it Department of Physics, Syracuse University}
\centerline{\it Syracuse NY 13244-1130, USA}
\smallskip
\centerline{and}
\smallskip
\centerline{Bill Spence}
\centerline{\it School of Physics, University of Melbourne}
\centerline{\it Parkville, 3052 Australia}
\bigskip
\centerline{\bf Abstract}
\noindent
We apply the inverse scattering method to the midi-superspace models
that are characterized by a two-parameter Abelian group of motions with
two spacelike Killing vectors. We present a formulation that simplifies
the construction of the soliton solutions of Belinski\v i and Zakharov.
Furthermore, it enables us to obtain the zero curvature formulation for these
models. Using this, and imposing periodic boundary conditions corresponding
to the Gowdy models when the spatial topology is a three torus $T ^3$, we show
that the equation of motion for the monodromy matrix is an evolution equation
of the Heisenberg type. Consequently, the eigenvalues of the monodromy matrix
are the generating functionals for the integrals of motion. Furthermore,
we utilise a suitable formulation of the transition matrix to obtain explicit
expressions for the integrals of motion. This involves recursion relations
which arise in solving an equation of Riccati type. In the case when the two
Killing vectors are hypersurface orthogonal the integrals of motion have a
particularly simple form.
\bigskip
\z
\eject}


\pageno=2
\noindent
{\bf 1. Introduction}
\medskip
\noindent
The Einstein field equations for space-times admitting a
two-dimensional Abelian group of isometries acting orthogonally
and transitively on non-null orbits are non-linear partial differential
equations in two variables. For timelike orbits the equations are elliptic,
whereas for spacelike orbits the equations are hyperbolic [1]. Although
the space-times admitting two commuting Killing vectors are not the most
general, they can represent interesting physical situations with
stationary axial symmetry, planar symmetry or cylindrical symmetry [2].

        Since the pioneering work by Geroch [3], it has been known
that the field equations in the stationary axisymmetric case admit an
infinite dimensional group of symmetry transformations. This result
has encouraged research in solution-generating methods, the idea
being that the complete class of solutions can be generated from a
particular solution, such as flat space [4]. Subsequently, several
solution-generating techniques have been developed, such as the
Kinnersley-Chitre transformations [5], the Hauser-Ernst formalism [6],
Harrison's B\" acklund transformations [7] and the Belinski\v i
and Zakharov inverse scattering method [8]. The relations between the
different approaches were discussed by Cosgrove [4], with Kitchingham [9]
adapting these methods to the hyperbolic case.

        Our paper uses the framework of the Belinski\v i and Zakharov
inverse scattering method, for the case when the field equations are
hyperbolic. The inverse scattering method is a powerful method for solving
certain systems of non-linear partial differential equations. The main step
in this formalism is to write down linear eigenvalue equations whose
integrability conditions are the given non-linear system. The methods
of functional analysis can be applied to generate new solutions of the
linear system from old, and hence new solutions of the original system
from old [10]. The particular solutions that can be generated are the
soliton solutions. The soliton solutions share a number of common properties
with classical particles, namely, they are localized solutions that
propagate energy, have a particular velocity of propagation and some
persistence of shape which is maintained even when two solitons collide [2].
As shown by Belinski\v i and Zakharov, the soliton transformation needs to
be generalized when applied in the context of the Einstein equations with a
two-parameter Abelian group of motions. The generalization is that the
stationary poles are substituted by the pole trajectories. We will present an
equivalent formulation of the linearized system. Similarly to
Belinski\v i and Zakharov [8], our linearized system is defined with the
use of two differential operators which involve derivatives with
respect to the complex parameter ${\l}$. We then define a new complex parameter
${\om}$, and show that this simplifies the linearized system. We also point out
the properties of the map between the two complex parameters
${\om} ( t , z , \l )$ and ${\l} ( t , z , \om )$. We then show
that our formulation yields soliton solutions equivalent to the original
solitons of Belinski\v i and Zakharov.

        We next construct the zero curvature formulation for the system.
The zero curvature formulation is an important characteristic of
integrable systems. A direct consequence of the zero curvature formulation,
for a given system, is the fact that, when periodic boundary conditions
are imposed, the equation of motion for the monodromy matrix is an evolution
equation of Heisenberg type [11]. Hence, the eigenvalues of the monodromy
matrix are conserved. In the context of the two spacelike Killing vector
reduction of general relativity, periodic boundary conditions amount
to the compactification of the $z$ direction into a circle. This then
corresponds to the Gowdy models when the spatial topology is a three torus
$T ^3$ [12].

        The next step is to obtain a suitable parameterization
for the transition matrix. To achieve this we have to solve a system of four
partial differential equations. This problem reduces to one of solving
two equations of Riccati type. We then look for solutions to the
Riccati equations in the form of power series in $({\l} \pm 1)$.
The solutions are given through recursion relations. For
integrable systems with fixed poles, the integrals of motion are then
given by the coefficients in the Laurent expansion of the eigenvalues of
the monodromy matrix around the poles $({\l} \pm 1)$.
However, in the case of the
Einstein field equations for space-times admitting a two-dimensional
Abelian group of isometries acting orthogonally and transitively, as we have
remarked earlier, the fixed poles are substituted by the pole trajectories,
\ie $( \, {\l} ( {\om} , t , z ) \pm 1 \, )$. Due to the fact that
$\l$ is time-dependent,
\ie $\partial _t  {\l} ( {\om} , t , z ) \ne 0$, we are not able to identify
the ``local'' integrals of motion as the coefficients in the expansion of the
eigenvalues of the transition matrix. Instead, ``local'' integrals of motion
are given as the eigenvalues of the transition matrix for fixed values
of the complex parameter ${\om}$ in the domain in which all the relevant
algebraic series converge uniformly. In the case when the Killing vectors
are hypersurface orthogonal the integrals of motion have a particularly
simple form.

        This paper is organized as follows. In section two we
formulate the inverse scattering method as applied
to the two Killing vector reduction of general relativity. This involves
the feature that the derivatives defining the first-order
formulation of the equations of motion also involve derivatives
with respect to the spectral parameter. We show how this may be dealt
with by defining a new complex parameter, and we discuss some properties of
this map. We then show how the approach of Belinski\v i and Zakharov
may be adapted to our case, and thus we obtain the soliton solutions.
In section three, we turn to the zero curvature representation of the
equations of motion, using this to define a transition matrix in the
usual way. The integrals of motion are then seen to be given in terms
of the related monodromy matrix. In section four, we utilise a suitable
formulation of the transition matrix to obtain explicit
expressions for the integrals of motion. This involves recursion relations
which arise in solving an equation of Riccati type. Finally, in section five
we present our conclusions.


\bigskip
\noindent
{\bf 2. The Inverse Scattering Method and the Soliton Solutions}
\medskip
\noindent
We will consider the midi-superspace models that are characterized
by the existence of a two-parameter Abelian group of motions with
two spacelike Killing vectors (the case when one Killing vector is
timelike and the other spacelike is similar and
we will not consider it separately). Let us choose coordinates
adapted to the action of the symmetry group so that the metric
assumes the following form [8]
$$
\eqalignno{ ds ^2 &=  - fdt ^2 + fdz ^2 + g _{ab} dx ^a dx ^b \, ,
                                                                & (2.1) \cr}
$$
where $a,b = 1,2$, $\{ x ^0, x ^1, x ^2, x ^3 \} =\{ t, x, y, z \}$,
$f$ is a positive function and $g _{ab}$ is a symmetric two-by-two matrix.
The function $f$ and the matrix $g _{ab}$ depend only on the co-ordinates
$\{ t,z \}$, or equivalently on the null co-ordinates $\{ {\x} , {\y} \} =
\{ \half ( z + t ), \half ( z - t ) \}$.
There is a freedom to perform the co-ordinate transformations
$$
\eqalignno{ \{{\x} , {\y} \} &\to \{ \tilde\xi(\xi),\tilde\eta(\eta) \}
                                            \, .                & (2.2) \cr}
$$
It is easy to see that the transformations (2.2) preserve both
the conformally
flat two-metric $f ( - dt ^2 + dz ^2)$ and the positivity of
the function $f$ if $\pde {\tilde\xi} {\x} \, \pde {\tilde\eta} {\y} > 0$.

The complete set of vacuum Einstein equations for the metric
(2.1) decomposes into two groups of equations [8]. The first group
determines the matrix $g _{ab}$ and can be written as a single
matrix equation
$$
\eqalignno{ \pde { \bigl ( \a \pde g {\x} \, {g ^{-1}} \bigr ) } {\y}
+ \pde { \bigl ( \a \pde g {\y} \, {g ^{-1}} \bigr ) } {\x} &= 0
                                         \, ,             & (2.3) \cr}
$$
where ${\a } ^2 = \det g$ and $\{ {\x} , {\y} \}$ are the null co-ordinates.
The second group of equations determines the function $f( \x , \y )$
in terms of a given solution of (2.3):
$$
\eqalignno{
\pde {(\ln f)} {\x} &= { \partial ^2 _{\x} {(\ln {\a})}\over{ \partial _{\x}
{(\ln {\a})} }} + { 1\over{ 4 {\a} \,  {\a} _{\x} }} Tr {\cal A} ^2 \, ,
                                                            & (2.4a) \cr
 \pde {(\ln f)} {\y} &= { \partial ^2 _{\y} {(\ln {\a})}\over{ \partial _{\y}
{(\ln {\a})} }} + { 1\over{ 4 {\a} \, {\a} _{\y} }} Tr {\cal B} ^2 \, ,
                                                             & (2.4b) \cr}
$$
where ${\a} _{\x} = \pde {\a} {\x}$, ${\a} _{\y} = \pde {\a} {\y}$ and
the matrices ${\cal A}$ and ${\cal B}$ are defined by
$$
\eqalignno{ {\cal A} = - {\a} \pde g {\x} \, g ^{-1},
\quad & \quad {\cal B} = {\a} \pde g {\y} \, g ^{-1}
                                                        \, . & (2.5) \cr}
$$
The dynamics of the system is thus essentially determined
by eqn. (2.3) and for this reason we will concentrate
on it in the following.

        Taking the trace of eqn. (2.3) and using the definition
for ${\a}$, we obtain
$$
\eqalignno {  {\a} _{\x \y} &= 0                \, .            & (2.6) \cr}
$$
The two independent solutions of this equation are
$$
\eqalignno{ {\a} &= a( {\x} ) + b ( {\y} ) \, ,                 & (2.7a) \cr
            {\b} &= a( {\x} ) - b ( {\y} ) \, .                 & (2.7b) \cr}
$$
Using the transformations (2.2), one can bring the functions $a({\x})$
and $b({\y})$ to a prescribed form. However, we will consider the general
form without specifying the functions $a({\x})$ and $b({\y})$ in advance.

Let us now consider the system of equations
$$
\eqalignno{ &\nabla _{\y} A + \nabla _{\x} B = 0  \, ,         & (2.8a) \cr
            &\pde A {\y} - \pde B {\x} + [ A , B ]  = 0
                                                  \, ,         & (2.8b) \cr}
$$
where $\nabla _{\x} {\z} = \pde {\z} {\x} + {\a} _{\x} {\a} ^{-1}$,
$\nabla _{\y} {\z} = \pde {\z} {\y} + {\a} _{\y} {\a} ^{-1}$ and
$[\ ,\ ]$ denotes the commutator in the Lie algebra of the group GL(2,R).
The general solution of the equation (2.8b) is of the form
$$
\eqalignno{ A = \pde{l} {\x} \, l^{-1} \, ,  \quad & \quad
            B = \pde{l} {\y} \, l^{-1} \, ,          & (2.9) \cr}
$$
where $l$ is an element of the group $GL(2,R)$. In addition we impose
the constraint ${\a } ^2 = \det l$. Eqn. (2.6) follows
from the trace of  eqn. (2.8a), once we substitute
eqn. (2.9) into (2.8a). However, we still have more degrees of freedom
in eqn. (2.8a) than in (2.3). We thus need to impose some additional
conditions in order to recover the correct number of degrees
of freedom. For this reason we impose the constraint
$l = l ^{\t}$, where $l^{\t}$ is the transpose of the matrix $l$.
It then follows, once the additional constraints are imposed,
that eqns. (2.8) are equivalent to eqn. (2.3).
Due to the constraint $l = l ^{\t}$, the matrices $A, B$ in eqn. (2.9)
must be taken to be only those which can be written in this form using a
{\it symmetric} matrix $l$. The sets of such matrices $A, B$ form
subsets (but not subgroups) of $GL(2,R)$.

The crucial step in the inverse scattering method
is to define the linearized system whose integrability conditions
are the equations of interest, in our case
eqns. (2.8). Following ref. [8], we first define the two
differential operators
$$
\eqalignno{ D _{\x} &= \pde {\z} {\x} - {{\a} _{\x}\over{\a} } \,
{{{\l} + 1}\over{ {\l} - 1 }} \, {\l} \pde {\z} {\l} ,  & (2.10a) \cr
            D _{\y} &= \pde {\z} {\y} - {{\a} _{\y}\over{\a} } \,
{{{\l} - 1}\over{ {\l} + 1 }} \, {\l} \pde {\z} {\l}, & (2.10b) \cr}
$$
where ${\l}$ is a complex parameter independent of the co-ordinates
$\{ {\x} , {\y} \}$. Notice that $D_{\x}$ and $D_{\y}$ are invariant
under the coordinate transformation
$\{ {\l} , {\x} , {\y} \} \to \{ {1\over \l} , {\x} , {\y} \}$.
It is straightforward to see that the
differential operators $D _{\x}$ and $D _{\y}$ commute since ${\a}$
satisfies the wave equation (2.6)
$$
\eqalignno{ [ D _{\x} , D _{\y} ] &= {{\a} _{\x \y}\over {\a}}
\bigr ( {{{\l} + 1}\over{ {\l} - 1 }}
- { {{\l} - 1}\over{ {\l} + 1 }}  \bigr )
{\l} \partial _{\l} = 0         \, .                  & (2.11) \cr}
$$
The next step is to consider the following linear system
$$
\eqalignno{ D _{\x} {\f} &= -{A\over{ {\l} - 1 }} {\f}
                                                \, ,  & (2.12a) \cr
            D _{\y} {\f} &= {B\over{ {\l} + 1 }} {\f}
                                                \, ,  & (2.12b) \cr}
$$
where ${\f} ( {\l} , {\x} , {\y} )$ is a complex matrix function,
and the real matrices $A$, $B$ and the real function ${\a}$
do not depend on the complex parameter ${\l}$. The integrability
conditions for the system (2.12) are eqns. (2.8). To prove
this we first apply the operator $D _{\y}$ to eqn. (2.12a) and
subtract from this the result of applying the operator $D _{\x}$
to eqn. (2.12b). The left-hand side vanishes using eqn. (2.11)
and the right-hand side is a rational function of ${\l}$
which vanishes if eqns. (2.8) are satisfied. In order to take
into account the additional constraints that we have imposed
we require that
$$
\eqalignno { \bar {\f} ( \bar {\l} ) &= {\f} ( {\l} )
                                        \, ,      & (2.13a) \cr
        g &= {\f} ( {1\over{\l} } ) {\f} ^{\t} ( {\l} )
                                        \, ,      & (2.13b) \cr}
$$
where $\bar {\l}$ is the complex conjugate to ${\l}$,
$g ({\x}, {\y})$ is a symmetric two-by-two matrix of functions,
whose determinant is $\a ^2$, and ${\f} (\l)$ satisfies eqns. (2.12).
The condition (2.13b) implies that $A$ and $B$ have the correct number
of degrees of freedom. To see this, we apply the $D_\xi$
operator to equation (2.13b). Using the fact that ${\f}$
satisfies eqn. (2.12a) we obtain
$$
\eqalignno { \partial _{\x} g &= {1\over{{\l} - 1}}
\Bigl ( {\l} \, A \, g - g \, A ^{\t} \Bigr )
                                          \, .       & (2.14a) \cr}
$$
Similarly, using (2.12b), we obtain
$$
\eqalignno { \partial _{\y} g &= {1\over{{\l} + 1}}
\Bigl ( {\l} \, B \, g + g \, B ^{\t} \Bigr )
                                          \, .       & (2.14b) \cr}
$$
Taking the transposes of the right-hand sides of equations (2.14),
we deduce that $A g = g A ^{\t}$ and $B g = g B ^{\t}$.
Consequently, $A$ and $B$ have the form

$$
\eqalignno{ A = \pde {g} {\x} \, g ^{-1},  \quad & \quad
            B = \pde { g} {\y} \, g ^{-1} \, .  & (2.15) \cr}
$$

The standard application of the inverse scattering method
to field theories in (1+1)-dimensions differs from the present
situation in that the linearized system
does not usually involve differentiation with respect to the spectral
parameter ${\l}$, as occurs here.
The present situation can be improved by defining
a new complex parameter $\omega$ by
$$
\eqalignno{\om &= {\half} \Bigl (
{ {\a} \over{\l} } + 2 {\b} + {\a \l} \Bigr )
                                    \, .                    & (2.16) \cr}
$$
A straightforward calculation shows that $D_\xi {\om} = D_\eta {\om} = 0$.
Furthermore, by making a co-ordinate transformation
$\{ {\l} , {\x} , {\y} \} \to \{ {\om} , {\x} , {\y} \}$
we reduce $D_\xi$ and $D_\eta$ to $\partial _{\x}$ and $\partial _{\y}$
respectively. In order to write the linearized system (2.12) in the
new co-ordinates we have to invert the relation (2.16), \ie\ we have
to know ${\l}$ as a function of  $( {\om} , {\x} , {\y} )$.
The inverse of (2.16) is not unique - we encounter the following two solutions
for ${\l}$:
$$
\eqalignno{ {\l} _{\pm} &= \Bigl ( {{\om} - {\b}\over {\a}} \Bigr ) {\pm}
\Bigl ( \, \Bigl ( {{\om} - {\b}\over {\a}} \Bigr ) ^2
- 1 \Bigr ) ^{\half} \, .  & (2.17) \cr}
$$
At this point, we have a choice of two different co-ordinate systems,
the first corresponding to ${\l} _{+}$ and the second to
${\l} _{-}$. We can now write the linearized system in the new
co-ordinates as
$$
\eqalignno{ \pde {\f} {\x} &= - { A\over{ {\l} _{\pm} - 1 } } {\f}
                                                \, ,  & (2.18a) \cr
            \pde {\f} {\y} &= { B\over{ {\l} _{\pm} + 1 } } {\f}
                                                \, ,  & (2.18b) \cr}
$$
where ${\f} ( {\om} , {\x} , {\y} )$ and ${\l} _{\pm}$ is given by (2.17).
It is straightforward to check that the integrability conditions for this
system are equations (2.8). To show this it is necessary to use equations
(2.7) and (2.17).

Let us point out some properties of the map (2.16).
The transformation $\l \to { 1 \over \l}$ leaves (2.16) invariant.
The $\l$ plane is mapped into the two-sheeted Riemann surface which
covers the entire $\l$ plane with the branch points at
$\om = \b \pm \a$. The map (2.16) takes the circle ${\abs {\l}} = \r$,
into the ellipse $C _{\r}$ given by the parametric equation
$$
\eqalignno { u &= {{\a}\over 2} \bigl ( {\r} + {1 \over{\r}} \bigr )
\cos \phi + \b \, ,                                     & (2.19a) \cr
             v &= {{\a}\over 2}  \bigl ( {\r} - {1 \over{\r}} \bigr )
\sin \phi \, ,                                          & (2.19b) \cr}
$$
where $\om = u + i v$ and $\phi$ is the phase of $\l$,
\ie $\l = \r e ^{i \phi}$. In particular, the image of the circle
${\abs {\l}} = 1$ is a closed interval on the real axis in the
$\om$ plane. From our previous discussion it follows that
the two formulations,
the first defined by $( {\l} , {\x} , {\y} )$ and the linearized system
(2.12), and the second defined by $( {\om} , {\x} , {\y} )$ and
the linearized system (2.18) are completely equivalent. In what follows
we will use both formulations.

        It is straightforward to obtain the soliton solutions
following the inverse scattering method [8]. We begin by setting
${\f} {\big |} _{{\l}={\pm}{\infty}} = \one$. Then, from eqn.
(2.13) it follows that $g ( {\x} , {\y} ) = {\f} ( 0 , {\x } , {\y} )$.
In the inverse scattering approach we start with a
given solution $g _0 ( {\x} , {\y} )$ of eqns.
(2.3). From eqn. (2.15) we then determine $A _0 ( {\x} , {\y} )$ and
$B _0 ( {\x} , {\y} )$. Substituting $A _0 ( {\x} , {\y} )$
and $B _0 ( {\x} , {\y} )$ into eqn. (2.12) and solving
the linearized system we obtain ${\f} _0 ( {\l} , {\x} , {\y} )$,
with $g _0 ( {\x} , {\y} ) = {\f} _0 ( 0 , {\x} , {\y} )$
\note{Integration of the linearized system is straightforward in
the case when the metric $g _0 ( {\x} , {\y} )$ is diagonal [13].
However, when $g _0 ( {\x} , {\y} )$ is non-diagonal
this step is non-trivial, an example of this being the case of the Bianchi
type II models [14,15].}
{}.
We now define ${\vf}$ as
$$
\eqalignno { {\f} ( {\l} , {\x } , {\y} ) &= {\vf} ( {\l} , {\x} , {\y} )
\, {\f} _0 ( {\l} , {\x} , {\y} )              \, .   &(2.20) \cr}
$$
The linearized eqns. for ${\vf}$, from eqns. (2.12), are
$$
\eqalignno{ D _{\x} {\vf} &= {1\over{ {\l} - 1 }} \Bigl (
- A {\vf} + {\vf} A _0 \Bigr )                 \, ,   & (2.21a) \cr
            D _{\y} {\vf} &= {1\over{ {\l} + 1 }} \Bigl (
  B {\vf} - {\vf} B _0 \Bigr )                 \, .   & (2.21b) \cr}
$$
We also have the constraint on ${\vf}$, from eqns. (2.13),
$$
\eqalignno { g ( {\x} , {\y} ) &= {\vf} ( {1\over {\l}} , {\x} , {\y} )
\, g _0 ( {\x} , {\y} ) \, {\vf} ^{\t} ( {\l} , {\x} , {\y} )
                                               \, .   &(2.22) \cr}
$$
Setting ${\vf} {\big |} _{{\l}={\pm}{\infty}} = \one$ in (2.22) it
follows that
$$
\eqalignno { g ( {\x} , {\y} ) &= {\vf} ( 0 , {\x} , {\y} )
\, g _0 ( {\x} , {\y} )
                                               \, .   &(2.23) \cr}
$$

        The soliton solutions are characterized by the points in
the ${\l}$ plane at which the determinant of ${\vf}$ is equal to zero and
${\vf} ^{-1}$ has a simple pole, and similarly, the points at which
the determinant of ${\vf} ^{-1}$ is equal to zero and ${\vf}$ has a simple
pole [8]. Thus, ${\vf}$ and ${\vf} ^{-1}$ are
rational matrix functions of ${\l}$ with a finite number of simple poles
$$
\eqalignno { {\vf} &= {\one} + \sum_{k=1}^N \Bigl (
{{\R} _k\over{ {\l} - {\mu} _k}} +
{{\bar {\R}} _k\over{ {\l} - {\bar {\mu} _k}}} \Bigr ) \, , &(2.24a) \cr
{\vf} ^{-1} &= {\one} + \sum_{k=1}^N \Bigl (
{{\So} _k\over{ {\l} - {\nu} _k}} +
{{\bar {\So}} _k\over{ {\l} - {\bar {\nu} _k}}} \Bigr ) \, , &(2.24b) \cr}
$$
where the matrices ${\R} _k$ and ${\So} _k$ do not depend on ${\l}$,
and ${\bar {\mu}} _k$ is the complex conjugate of ${\mu} _k$, and
${\bar {\nu}} _k$ of ${\nu} _k$. From eqn. (2.24a), it is easy to
check that the condition ${\vf} {\big |} _{{\l}={\pm}{\infty}} = \one$
is satisfied. Also, from eqn. (2.22) it follows that the positions of
the poles of
${\vf}$ and the poles of ${\vf} ^{-1}$ are related by
${\mu} _k {\nu} _k= 1$, $k = 1,....,N$. This implies, using eqn. (2.16),
that ${\om} _k = {\half} \Bigl ( { {\a} \over{{\mu} _k} }
+ 2 {\b} + {\a {\mu} _k} \Bigr ) = {\half} \Bigl ( { {\a} \over{{\nu} _k} }
+ 2 {\b} + {\a {\nu} _k} \Bigr )$, $k = 1, .... , N$.

        The next step is to rewrite eqns. (2.21) in the more
convenient form
$$
\eqalignno{ - {A\over{ {\l} - 1 }} &=  \Bigl ( D _{\x} {\vf} \Bigr )
{\vf} ^{-1} - {\vf} {A _0\over{ {\l} - 1 }} {\vf} ^{-1}
                                                \, ,   & (2.25a) \cr
              {B\over{ {\l} + 1 }} &= \Bigl ( D _{\y} {\vf} \Bigr )
{\vf} ^{-1} + {\vf} {B _0\over{ {\l} + 1 }} {\vf} ^{-1}
                                                \, .   & (2.25b) \cr}
$$
Substituting eqn. (2.24) into (2.25) and setting the residuals at the poles
${\l}={\mu} _k$ of the right-hand-side equal to zero, we obtain a set
of $N$ first order equations for the matrices ${\R} _k$, $k = 1, .... , N$.
Similarly, from the identity ${\vf} \, {\vf} ^{-1} = \one $ we obtain
the system of $N$ algebraic equations
$$
\eqalignno { {\R} _k {\vf} ^{-1} ( {\mu} _k ) &= 0
                                                \, ,   &(2.26a) \cr}
$$
and from eqn. (2.22) we obtain another system of $N$ algebraic equations
$$
\eqalignno { {\R} _k \, g _0 \, \Bigl ( {\one} + \sum _{l=1}^N
\Bigl ( {{{\R} _l} ^{\t}\over{ {{\nu} _k} - {\mu} _l}} +
{{{\bar {\R}} _l} ^{\t}\over{ {{\nu} _k} - {\bar {\mu} _l}}} \Bigr )
\, \Bigr ) &= 0                                 \, ,    &(2.26b) \cr}
$$
where we have used the identity ${\mu} _k {\nu} _k = 1$.
We also have similar equations for the matrices ${\bar {\R}} _k$ and
the poles ${\l} = {\bar {\mu}} _k$, $k= 1 , ... , N$.
This yields the $n$-soliton solutions equivalent to those of
Zakharov and Belinski\v i, [8]. In addition, equations (2.4),
which determine the function $f$, can then be integrated explicitly
following ref. [8,16]. We will not present these results here.
Instead, we will turn to the study of the zero curvature formulation and
the integrals of motion.

\bigskip

\noindent
{\bf 3. The Zero Curvature Formulation and The Integrals of Motion}
\medskip
\noindent
We now construct the zero curvature formulation for equations
(2.8) and, using the techniques of ref. [11] in the case when the $z$
direction is compactified into a circle $S ^1$, obtain the
generating functional for the integrals of motion. Our first step
is to perform a co-ordinate transformation $\{ {\x} , {\y} \} \to
\{ t , z \}$ and define $U ( t , z , {\l} )$ and $V ( t , z , {\l} )$ by
$$
\eqalignno { U ( t , z , {\l} ) &= {\half}
\Bigl (- { A\over{ {\l} - 1 } } + { B\over{ {\l} + 1 } } \Bigr )
                                                \, , &(3.1a) \cr
             V ( t , z , {\l} ) &= {\half}
\Bigl (- { A\over{ {\l} - 1 } } - { B\over{ {\l} + 1 } } \Bigr )
                                                \, . &(3.1b) \cr}
$$
The equation of motion for $U$ and $V$ is
$$
\eqalignno { D _t U - D _z V +
{\bigl [} U , V {\bigr ]} &= 0   \, ,  &(3.2) \cr}
$$
where $D _t = {\half} \Bigl ( D _{\x} - D _{\y} \Bigr )$, $D _z = {\half}
\Bigl ( D _{\x} + D _{\y} \Bigr )$, with $D _{\x}$ and $D _{\y}$ given by
eqn. (2.10). Substituting eqn. (3.1) into (3.2), it is straightforward to
show that eqn. (3.2) is equivalent to eqns. (2.8).

        The fields $U ( t , z , {\om} )$ and $V ( t , z , {\om} )$
are expressed, in the co-ordinates $\{ t , z , \om \}$, by the formulae
that can be obtained from (3.1) by the substitution $\l \to \l _+
( t , z , {\om} )$, where $\l _+ ( t , z , {\om} )$ is defined in (2.17)
\note{From hereon we will restrict ourselves to the plus sign
in eqn. (2.17).}
. Consequently, the equation of motion in this co-ordinate system is
the zero curvature equation, \ie\ eqn. (3.2) with the substitution
$D _t \to \partial _t$ and $D _z \to \partial _z$. We will consider
the case when periodic boundary conditions are imposed
$$
\eqalignno {  U ( t , z+2L , {\om} ) &=
              U ( t , z , {\om} )           \, , &(3.3a) \cr
              V ( t , z+2L , {\om} ) &=
              V ( t , z , {\om} )           \, . &(3.3b) \cr}
$$
Notice that we impose periodic boundary conditions on all
fields, including ${\a}$ and ${\b}$. These boundary conditions
correspond to compactifying the $z$ direction into a circle $S ^1$.
For example, this corresponds to the Gowdy models
when the spatial topology is a three torus $T ^3$ [12].
We consider the transition matrix
$$
\eqalignno { T ( t, z _1 , z _0, {\om}) &= {\Pe} \exp
\int _{z _0}^{z _1} U ( t , z , {\om} ) \, dz  \, . &(3.4) \cr}
$$
The transition matrix satisfies
$$
\eqalignno { \partial _{z _1} T ( t, z _1 , z _0, {\om} ) &=
U ( t , z _1 , {\om} ) \, T ( t, z _1 , z _0 , {\om})
                                              \, , &(3.5) \cr}
$$
with the condition
$$
\eqalignno { T ( t, z _1 , z _0 , {\om} )
{\Big |} _{z_0=z_1} &= {\one}                 \, . &(3.6) \cr}
$$
We apply $\partial _t$ to eqn. (3.5), obtaining
$$
\eqalignno { \partial _t (\partial _z T) &= \partial _t (U) \, T
+ U \partial _t (T)
                                              \, . &(3.7) \cr}
$$
Using the equations of motion for $\{ U , V \}$, the zero curvature
equation (3.7) can be written
$$
\eqalignno { \partial _z \bigl ( \partial _t T - V T \bigr ) &=
U \bigl ( \partial _t T - V T \bigr )                \, . &(3.8) \cr}
$$
It follows from (3.5) that
$$
\eqalignno { \partial _t T (t,z_1,z_0,{\om}) &=
V (t,z_1,{\om}) \, T (t,z_1,z_0,{\om})           \cr
&+ T (t,z_1,z_0,{\om}) \, C (t,z _0,{\om})
                                           \, , &(3.9) \cr}
$$
and, using the condition (3.6), we get $C (t,z _0,{\om}) =
- V (t , z _0 , {\om})$. Thus the equation of motion for the
transition matrix is
$$
\eqalignno { \partial _t T (t,z_1,z_0,{\om}) &=
V (t,z_1,{\om}) \, T (t,z_1,z_0,{\om})           \cr
&- T (t,z_1,z_0,{\om}) \, V (t,z _0,{\om})
                                           \, , &(3.10) \cr}
$$

        We now define the monodromy matrix to be the transition matrix
along the fundamental domain $-L \leq z \leq L$, \ie
$$
\eqalignno { T _L ( t , {\om}) &= T ( t , L , -L , {\om} )
                                                \, . &(3.11) \cr}
$$
{}From eqn. (3.10), using the periodic boundary conditions (3.3), it follows
that the equation of motion for the monodromy matrix $T _L ( t , {\om} )$
is an evolution equation of Heisenberg type
$$
\eqalignno { \partial _t T _L (t , {\om}) &=
{\bigl [} V (t,L,{\om}) , T _L ( t , {\om}) {\bigr ]}
                                        \, .    &(3.12) \cr}
$$
This implies that the eigenvalues of the monodromy matrix
$T _L ( t , {\om} )$ are conserved, or equivalently
$$
\eqalignno { \partial _t {\Tr} \, T _L (t , {\om}) &= 0
                                        \, ,    &(3.13a) \cr
             \partial _t {\Tr} \, \bigl ( T _L \bigr ) ^2
(t,{\om}) &= 0                           \, .    &(3.13b) \cr}
$$
Our conclusion is that the functions
$$
\eqalignno { E _L ({\om}) &= {\Tr} \, T _L (t,{\om})
                            \, ,       &(3.14a) \cr
             F _L ({\om}) &= {\Tr} \, \bigl ( T _L \bigr ) ^2 (t,{\om})
                            \, ,       &(3.14b) \cr}
$$
are the generating functions for the conservation laws. This result
is a direct consequence of the zero curvature formulation and the
periodic boundary conditions as shown above. Similar
results are known for most if not all of the dynamical systems that
admit a zero curvature formulation [11]. The standard method
to obtain the explicit expressions for the integrals of motion involves
solving the equations of Riccati type. The integrals of motion are
then identified as the coefficients in the Laurent expansions of the
generating functions. However, as we will show in the next section,
in our case the generating functions (3.14) depend on the complex parameter
$\om$ through ${\l} _+ ( t , z , \om )$. Consequently, we cannot identify
the integrals of motion as the coefficients in the expansion of the
generating functions. Instead, our integrals of motion are given by
algebraic series, for those values of $\om$ for which the series
converge uniformly.

\bigskip

\noindent
{\bf 4. The Integrals of Motion}
\medskip
\noindent
We can now obtain explicit expressions for the integrals
of motion. We first write the transition matrix in the form
$$
\eqalignno { T ( t, z _1 , z _0, {\om}) &= \Bigl (
{\one} + W ( t, z _1 , {\om} ) \Bigr )
e ^{ Z ( t, z _1 , z _0, {\om} )}
\Bigl ( {\one} + W ( t, z _0 , {\om} ) \Bigr ) ^{-1}
                                   \, , & (4.1) \cr}
$$
where ${\one}$ is the identity two-by-two matrix,
$W ( t, z , {\om} )$ is an off-diagonal matrix and
$Z ( t, z _1 , z _0, {\om} )$ is a diagonal matrix [11].
Substituting eqn. (4.1) into (3.5) we obtain the following system
of equations
$$
\eqalignno { \partial _{z _1} Z ^{(1)} - {\half} \Bigl ( U ^{(3)} W ^{(4)}
+ U ^{(4)} W ^{(3)} \Bigr ) - U ^{(1)} &=0 \, , &\quad (4.2a) \cr
\partial _{z _1} Z ^{(2)} - {\half} \Bigl ( U ^{(3)} W ^{(4)}
- U ^{(4)} W ^{(3)} \Bigr ) - U ^{(2)} &=0 \, , &\quad (4.2b) \cr
\partial _{z _1} W ^{(3)} + W ^{(3)}\Bigl( \partial _{z _1} Z ^{(1)}
- \partial _{z _1} Z ^{(2)} \Bigr ) - \Bigl(U ^{(1)}+ U ^{(2)} \Bigr )
W ^{(3)} - U ^{(3)} &=0                         \, , &\quad (4.2c) \cr
\partial _{z _1} W ^{(4)} + W ^{(4)}\Bigl( \partial _{z _1} Z ^{(1)}
+ \partial _{z _1} Z ^{(2)} \Bigr )   - \Bigl(U ^{(1)} - U ^{(2)}\Bigr )
W ^{(4)} - U ^{(4)} &=0                         \, . &\quad (4.2d) \cr}
$$
Here we have used the notation $Z=Z ^{(1)} \, {\t} _{(1)} + Z ^{(2)}
\, {\t} _{(2)}$,  $W = W ^{(3)} {\t} _{(3)} + W ^{(4)} {\t} _{(4)}$ and
$U  = \sum_{i=1}^{4} U ^{(i)}  \, {\t} _{(i)}$. Our
choice for a basis in the Lie algebra of the group
GL(2,R) is ${\t} _{(1)} = {\one}$,
${\t} _{(2)} = \bigl ( {1\atop 0} {0\atop -1} \bigr )$,
${\t} _{(3)} = \bigl ( {0\atop 0} {1\atop 0} \bigr )$ and
${\t} _{(4)} = \bigl ( {0\atop 1} {0\atop 0} \bigr )$.

Substituting eqns. (4.2a,b) into eqns. (4.2c,d), we
obtain the system of equations that determines
$W ( t, z _1 , {\om} )$
$$
\eqalignno { \partial _{z _1} W ^{(3)} + U ^{(4)} \Bigl ( W ^{(3)}
\Bigr ) ^2 - 2 U ^{(2)} W ^{(3)} - U ^{(3)} &=0       \, , & (4.3a) \cr
\partial _{z _1}  W ^{(4)} + U ^{(3)} \Bigl ( W ^{(4)} \Bigr ) ^2
+ 2 U ^{(2)} W ^{(4)} - U ^{(4)} &=0       \, , & (4.3b) \cr}
$$
with periodic boundary conditions on $W ( t, z _1 , {\om} )$, \ie
$W ( t, z _1 , {\om} ) = W ( t, z _1 + 2L, {\om} )$.
Once we have solved eqns. (4.3) for $W ( t, z_1 , {\om} )$,
eqns. (4.2a,b), together with the boundary condition
$Z ( t, z _1 , z _0, {\om} ) {\big |} _{z _1 = z _0} = 0$,
determine $Z ( t, z _1 , z _0, {\om} )$:
$$
\eqalignno { Z ^{(1)} ( t, z _1 , z _0, {\om} ) &=
\int _{z_0}^{z_1} \! dz \Bigl (
U ^{(1)} + {\half} \Bigl ( U ^{(3)} W ^{(4)}
+ U ^{(4)} W ^{(3)} \Bigr ) \, \Bigr )    \, , & (4.4a) \cr
Z ^{(2)} ( t, z _1 , z _0, {\om} )
&= \int _{z_0}^{z_1} \! dz \Bigl (
U ^{(2)} + {\half} \Bigl ( U ^{(3)} W ^{(4)}
- U ^{(4)} W ^{(3)} \Bigr ) \, \Bigr )     \, . & (4.4b) \cr}
$$

        The main problem is thus to obtain the solutions to the system
(4.3). These equations are of Riccati
type. Given an equation of Riccati type with arbitrary
coefficients together with a particular solution, it is possible
to reduce the equation to a linear first order system [17]. However, we
do not have particular solutions to equations (4.3), so
we need a different approach. We make a coordinate transformation
$\{ t , z _1 , {\om} \} \to \{ t , z_1 , {\l} \}$ and as a result
we obtain the equations (4.3) in the form
$$
\eqalignno { D _{z _1} W ^{(3)} + U ^{(4)} \Bigl ( W ^{(3)}
\Bigr ) ^2 - 2 U ^{(2)} W ^{(3)} - U ^{(3)} &=0       \, , & (4.5a) \cr
D _{z _1}  W ^{(4)} + U ^{(3)} \Bigl ( W ^{(4)} \Bigr ) ^2
+ 2 U ^{(2)} W ^{(4)} - U ^{(4)} &=0       \, , & (4.5b) \cr}
$$
We now expand the fields $W ^{(3)}$ and $W ^{(4)}$ as power series in
$({\l} - 1 )$
$$
\eqalignno { W ^{(3)} = \sum _{n=0}^{\infty}
W ^{(3)}_n ({\l} - 1 ) ^n                      \, ,& \quad
W ^{(4)} = \sum _{n=0}^{\infty} W ^{(4)}_n ({\l} - 1 ) ^n
                                                \, . &(4.6) \cr}
$$
Substituting the first equation of (4.6) and (3.1a) into (4.5a) and
using the expansion ${1\over {1+{\l}}} = {\half} \Sigma _{n=0}^{\infty}
\bigl ( - {\half} \bigr ) ^n ( {{\l} - 1} ) ^n$
we obtain the recursion relation ($N = 0,1,2,\dots$)
$$
\eqalignno { &(N + 2) \, {{\a} _{\x}\over {\a}} W _{N+2}^{(3)} =
\partial _{z_1} W _N^{(3)} - {A ^{(4)}\over 2}
\sum_{n=0}^{N+1} W _n^{(3)} W _{N+1-n}^{(3)}
+ {B ^{(4)}\over 4} \sum _{m=0} ^N \Bigl ( - {1\over 2} \Bigr ) ^m \times
                                                                      \cr
&\sum _{n=0}^{N-m} W _n^{(3)} W _{N-m-n}^{(3)}
- \Bigl ( {3\over 2}
(N+1) {{\a} _{\x}\over {\a}} - A ^{(2)} \Bigr ) W _{N+1} ^{(3)}
- {N\over 2} \, {{\a} _{\x}\over {\a}} \, W _{N}^{(3)}
- {1\over 4} \sum _{n=0}^N \Bigl ( - {\half} \Bigr ) ^n
\times                                                                \cr
&\Bigl ( {{\a} _{\y}\over {\a}} \Bigl ( N - n \Bigr ) + 2 B ^{(2)} \Bigr )
W _{N-n} ^{(3)} - {{\a} _{\y}\over 4{\a}} \sum _{n=1}^N
\Bigl ( - {\half} \Bigr ) ^{n-1} \Bigl ( N - n \Bigr ) W ^{(3)}_{N-n} +
{B ^{(3)}\over 2} \Bigl ( - {\half} \Bigr ) ^{N+1}
                                            \,   &(4.7) \cr}
$$
together with
$$
\eqalignno { {{\a} _{\x}\over {\a}} W _1^{(3)} &=
-{A ^{(4)}\over 2} \Bigl ( W ^{(3)}_0 \Bigr ) ^2
+ A ^{(2)} W ^{(3)}_0 + {A ^{(3)}\over 2}       \, . &(4.8) \cr}
$$
Notice that $W _0^{(3)}$ is arbitrary, and for every choice of
$W _0^{(3)}$ we have a different solution.
{}From eqn. (4.5b), we obtain similar equations determining the $W^{(4)}_n$,
which may be obtained from eqns. (4.7) and (4.8) by the
replacements $W^{(3)}_n \rightarrow W^{(4)}_n$,
$A^{(4)}\leftrightarrow A^{(3)}$, $B^{(4)}\leftrightarrow B^{(3)}$,
$A^{(2)}\rightarrow -A^{(2)}$ and $B^{(2)}\rightarrow -B^{(2)}$.

        In this way we have obtained the solutions to equations
(4.5) in an open neighborhood of the point $\l = 1$. Our next step is to
perform a coordinate transformation $\{ t , z_1 , {\l} \} \to
\{ t , z _1 , {\om} \}$. Then
$W = W ( t , z , \l _+ ( t , z , \om ))$ becomes a function of
$t$, $z$ and $\om$, and eqn. (4.6) becomes an
algebraic series in powers of $(\l _+ ( t , z , \om ) - 1 )$.
Substituting this expansion and the expansion ${1\over {1 + {\l} _+}} = {\half}
\sum _{n=0}^{\infty} \bigl ( - {\half} \bigr ) ^n ( {\l} _+ - 1 ) ^n$
into (4.4) we obtain
$$
\eqalignno { Z ^{(1)} ( t , z _1 , z _0 , \om ) &= \int _{z_0}^{z_1} \! dz
\Bigl ( - {\half} {1\over {{\l} _+ - 1}}   \Bigl ( A ^{(1)} + {\half}
\Bigl ( A ^{(3)} W ^{(4)}_0 + A ^{(4)} W ^{(3)}_0 \Bigr ) \, \Bigr ) \cr
&+ {1\over 4} \sum _{n=0}^{\infty} \Bigl ( B ^{(1)} \Bigl ( - {\half}
\Bigr ) ^n - \Bigl ( A ^{(3)} W ^{(4)}_{n+1} + A ^{(4)} W ^{(3)}_{n+1}
\Bigr )                                                         \cr
&+ {\half}  \sum _{m=0}^n  \, \Bigl ( - {\half} \Bigr ) ^m  \,
\Bigl ( B ^{(3)} W ^{(4)} _{n-m} + B ^{(4)} W ^{(3)} _{n-m} \Bigr )
 \, \Bigr ) ({\l}_+ - 1 ) ^n \Bigr )              \, , &(4.9a) \cr
Z ^{(2)} ( t , z _1 , z _0 , \om ) &= \int _{z_0}^{z_1} \! dz \Bigl (
- {\half} {1\over {{\l} _+ - 1}}   \Bigl ( A ^{(2)} + {\half}
\Bigl ( A ^{(3)} W ^{(4)}_0 - A ^{(4)} W ^{(3)}_0 \Bigr ) \, \Bigr ) \cr
&+ {1\over 4} \sum _{n=0}^{\infty} \Bigl ( B ^{(2)} \Bigl ( - {\half}
\Bigr ) ^n - \Bigl ( A ^{(3)} W ^{(4)}_{n+1} - A ^{(4)} W ^{(3)}_{n+1}
\Bigr )                                                         \cr
&+ {\half}  \sum _{m=0}^n  \, \Bigl ( - {\half} \Bigr ) ^m  \,
\Bigl ( B ^{(3)} W ^{(4)} _{n-m} - B ^{(4)} W ^{(3)} _{n-m} \Bigr )
 \, \Bigr ) ({\l} _+ - 1 ) ^n \Bigr )              \, . &(4.9b) \cr}
$$
where $W^{(3)}_n$ and $W^{(4)}_n$ are determined from eqns. (4.7) and (4.8).
Thus we have obtained the functions
$Z ^{(i)} ( t , z _1 , z _0 , \om )$, defined on open neighborhoods
of the hypersurface $ \l _+ ( t , z , \om ) = 1$ where the relevant
algebraic series converge uniformly.

        We now proceed to construct the functions
$Z ^{(i)} ( t , z _1 , z _0 , \om )$, defined on open
neighborhoods of the hypersurface $\l _+ ( t , z , \om ) = -1$. Our first
step is to construct the solution to the equation (4.5) in an open
neighborhood of the point $\l = -1$. We expand the fields $W ^{(3)}$
and $W ^{(4)}$ in powers of $({\l} + 1 )$
$$
\eqalignno { W ^{(3)} = \sum _{n=0}^{\infty}
W ^{(3)}_n ({\l} + 1 ) ^n                      \, ,& \quad
W ^{(4)} = \sum _{n=0}^{\infty} W ^{(4)}_n ({\l} + 1 ) ^n
                                                \, , &(4.10) \cr}
$$
and the expansion ${1\over {{\l}-1}} = - {\half}
\sum _{n=0}^{\infty} \bigl ( {{{\l} + 1}\over 2} \bigr ) ^n$
leads to the following recursion relation ($N = 0,1,2,\dots$)
$$
\eqalignno { &(N + 2) \, {{\a} _{\y}\over {\a}} W _{N+2}^{(3)} =
\partial _{z_1} W _N^{(3)} + {B ^{(4)}\over 2}
\sum _{n=0}^{N+1} W _n^{(3)} W _{N+1-n}^{(3)}
+{A ^{(4)}\over 4} \sum _{m=0} ^N {1\over 2 ^m}   \times        \cr
& \sum_{n=0}^{N-m} W _n^{(3)} W _{N-m-n}^{(3)} + \Bigl ( {3\over 2}
(N+1) {{\a} _{\y}\over {\a}} - B ^{(2)} \Bigr ) W _{N+1} ^{(3)}
- {N\over 2} \, {{\a} _{\y}\over {\a}} W _{N}^{(3)}
- {1\over 4} \sum_{n=0}^N {1\over 2^n}   \times                 \cr
&\Bigl ({{\a} _{\x}\over {\a}} \,  \Bigl ( N - n \Bigr )
+ 2 A ^{(2)} \Bigr ) W _{N-n} ^{(3)} + {{\a} _{\x}\over 4{\a}}
\sum_{n=1}^N {1\over 2^{n-1}} \Bigl ( N - n \Bigr ) W _{N-n} ^{(3)}
-{A ^{(3)}\over 2^{N+2}}
                                               \, , &(4.11) \cr}
$$
together with
$$
\eqalignno { {{\a} _{\y}\over {\a}} W _1^{(3)} &= {B ^{(4)}\over 2}
\Bigl ( W _0^{(3)} \Bigr ) ^2 - B ^{(2)} W _0^{(3)} - {\half} B ^{(3)}
                                                \, . &(4.12) \cr}
$$
As with eqn. (4.8), $W _0^{(3)}$ is arbitrary, and for
different choices of $W _0^{(3)}$ we have different solutions.
{}From (4.5b) and (4.10) we can obtain similar equations determining
the $W^{(4)}_n$, by the following replacements in those equations:
$W^{(3)}_n \rightarrow W^{(4)}_n$, $A^{(4)}\leftrightarrow A^{(3)}$,
$B^{(4)}\leftrightarrow B^{(3)}$, $A^{(2)}\rightarrow -A^{(2)}$ and
$B^{(2)}\rightarrow -B^{(2)}$.

        We now perform the coordinate transformation $\{ t , z , \l \}
\to \{ t , z , \om \}$. Then (4.10) becomes
an algebraic series in powers of $(\l _+ + 1)$.
Using the expansions (4.10,11) and ${1\over {{\l}_+ -1}} = - {\half}
\sum _{n=0}^{\infty} \bigl ( {{{\l} _+ + 1}\over 2} \bigr ) ^n$
we obtain
$$
\eqalignno { Z ^{(1)} ( t , z _1 , z _0 , \om ) &= \int _{z_0}^{z_1} \! dz
\Bigl ( {\half} {1\over {{\l} _+ + 1}}   \Bigl ( B ^{(1)} + {\half}
\Bigl ( B ^{(3)} W ^{(4)}_0 + B ^{(4)} W ^{(3)}_0 \Bigr ) \, \Bigr ) \cr
&+ {1\over 4} \sum _{n=0}^{\infty} \Bigl ( A ^{(1)} {1\over 2^n}
+ \Bigl ( B ^{(3)} W ^{(4)}_{n+1} + B ^{(4)} W ^{(3)}_{n+1}
\Bigr )                                                         \cr
&+ \sum _{m=0}^n  \, {1\over 2^{n+1}}  \,
\Bigl ( A ^{(3)} W ^{(4)} _{n-m} + A ^{(4)} W ^{(3)} _{n-m} \Bigr )
 \, \Bigr ) ({\l}_+ + 1 ) ^n \Bigr )              \, , &(4.13a) \cr
Z ^{(2)} ( t , z _1 , z _0 , \om ) &= \int _{z_0}^{z_1} \! dz
\Bigl ( {\half} {1\over {{\l} _+ + 1}}   \Bigl ( B ^{(2)} + {\half}
\Bigl ( B ^{(3)} W ^{(4)}_0 - B ^{(4)} W ^{(3)}_0 \Bigr ) \, \Bigr ) \cr
&+ {1\over 4} \sum _{n=0}^{\infty} \Bigl ( A ^{(2)} {1\over 2^n}
+ \Bigl ( B ^{(3)} W ^{(4)}_{n+1} - B ^{(4)} W ^{(3)}_{n+1}
\Bigr )                                                         \cr
&+ \sum _{m=0}^n  \, {1\over 2^{n+1}}  \,
\Bigl ( A ^{(3)} W ^{(4)} _{n-m} - A ^{(4)} W ^{(3)} _{n-m} \Bigr )
 \, \Bigr ) ({\l}_+ + 1 ) ^n \Bigr )              \, , &(4.13b) \cr}
$$
where $W^{(3)}_n$ and $W^{(4)}_n$ are determined from the recursion relations
derived for them above. We have thus determined the functions
$Z ^{(i)} ( t , z _1 , z _0 , \om )$ in an open neighbourhood
of the hypersurface $\l _+ ( t , z , \om ) = -1$ where
the relevant algebraic series converge uniformly.

       Since we know (locally) the functions
$Z ^{(i)} ( t , z _1 , z _0 , \om )$, we can obtain expressions
for the integrals of motion of our system. Using (4.1) and the fact that
$W ( t , z , {\om} ) = W ( t , z + 2L , {\om} )$ it follows that
$$
\eqalignno { E _L ({\om}) &= {\Tr} \, T _L (t,{\om})
= {\Tr} \, e ^{ Z ( L , -L , {\om} )}     \, ,       &(4.14a) \cr
             F _L ({\om}) &= {\Tr} \, \bigl ( T _L \bigr ) ^2 (t,{\om})
= {\Tr} \, e ^{ 2 Z ( L , -L , {\om} )}   \, .       &(4.14b) \cr}
$$
{}From (4.14), after a straightforward calculation, it follows that
$$
\eqalignno { Z ^{(1)} ( L , -L , {\om} ) &= {\half} \ln \Bigl (
{{{E _L} ^2 (\om) - F _L (\om)}\over 2} \Bigr )     \, , &(4.15a) \cr
             Z ^{(2)} ( L , -L , {\om} ) &= {\half} \cosh ^{-1} \Bigl (
{F _L (\om)\over {{E _L} ^2 (\om) - F _L (\om)}} \Bigr )
                                              \, .    &(4.15b) \cr}
$$
Consequently, from (3.13) and (4.15), we have
$$
\eqalignno { \partial _t Z ^{(i)} ( L , -L , {\om} ) &= 0
				               \, .   &(4.16) \cr}
$$
The conclusion is that the $Z ^{(i)} ( L , -L , {\om} )$ are the integrals
of motion for every fixed value of $\om$ which belongs to the domain
in which the relevant series converge uniformly.

	Our final remark is on the case when the two Killing vectors
are hypersurface orthogonal. In our formulation, this case corresponds to
the vanishing of the off-diagonal matrix $W$ in (4.1). The transition matrix
$T$ is then diagonal. Consequently, our integrals of motion have a
particularly simple form
$$
\eqalignno { Z ^{(1)} ( L , -L , {\om} ) &=
\int _{-L}^{L} \! dz \, U ^{(1)}   \, ,  	& (4.17a) \cr
Z ^{(2)} ( L , -L , {\om} ) &=
\int _{-L}^{L} \! dz  \, U ^{(2)}   \, , 	& (4.17b) \cr}
$$
with
$$
\eqalignno {
U ^{(1)} &= {1\over 2} \Bigl ( - {{\a} _{\x} \over {{\l} _+ - 1}}
+ {{\a} _{\y} \over {{\l} _+ + 1}} \Bigr )
				    \, , 	& (4.18a) \cr
 U ^{(2)} &= {1\over 2} \Bigl ( - {{\g} _{\x} \over {{\l} _+ - 1}}
  + {{\g} _{\y} \over {{\l} _+ + 1} } \Bigr )
				    \, . 	& (4.18b) \cr}
$$
To obtain (4.18) we have used equations (2.15) together with
the parametrization
$$
\eqalignno { g &= e^{{\a} {\t} _{(1)} + {\g} {\t} _{(2)}}
				    \, . 	& (4.19) \cr}
$$

	Work on the geometrical interpretation of the integrals of motion
$Z ^{(i)} ( L , - L , \om )$ and the boundary conditions that correspond
to gravitational plane waves and gravitational cylindrical waves is
in progress [18].


\bigskip
\noindent
{\bf 5. Conclusions}
\medskip
\noindent
In this paper we have formulated the inverse scattering method as applied
to the midi-superspace models characterized by a
two-parameter Abelian group of motions with two spacelike Killing vectors.
The application of the inverse scattering method to this model involved
the feature that the derivatives defining the first-order formulation of
the equations of motion also involved derivatives with respect to the
spectral parameter $\l$. We dealt with this by introducing a new
spectral parameter $\om$. We also discussed the properties of the map
between the two complex parameters $\om ( t , z , \l )$ and
$\l ( t , z , \om )$. Then we showed how the approach of
Belinski\v i and Zakharov could be adapted to our case, and hence found
the soliton solutions.

        Our next step was to obtain the zero curvature representation
of the Einstein field equations for space-times admitting a two-dimensional
Abelian group of isometries, in the case when the orbits are spacelike.
Then, using periodic boundary conditions which in the present context
correspond to the three torus Gowdy models,
we showed how the zero curvature formulation implies that the
equations of motion for the transition matrix are of Heisenberg type.
Consequently the eigenvalues of the transition matrix are conserved.

        To obtain explicit expressions for the integrals of motion we had
to solve a system of four partial differential equations. This problem
was reduced to the problem of solving two equations of Riccati
type. The solutions of these equations were given through recursion
relations. The final results are the integrals of motion which are given
as powers series in $( \, {\l} _+ ( t , z , \om ) \pm 1 \, )$.
In the case when the two Killing vectors are hypersurface orthogonal
the integrals of motion have a particularly simple form.

        Future research will include the elaboration of the geometrical
interpretation of the integrals of motion that we have obtained, as well as
an investigation of different boundary conditions, such as those
corresponding to planar symmetry and cylindrical symmetry [18].
The problem of quantization of the soliton solutions on the Bianchi
type II background is under investigation and we will report our results
shortly [15]. Finally, our main goal is to obtain the quantum theory
for the two Killing vector reduction of general relativity [19].


\bigskip
\noindent
{\bf 6. Acknowledgments}
\medskip
\noindent
We would like to thank A. Ashtekar, N. Kalogeropoulos, G. Mena-Marug\' an,
\break
R.S. Rajeev, L. Smolin, R. Sorkin, and G. Stephens for discussions.
N.M. is very grateful to A. Ashtekar for financial support and
the warm hospitality of the Relativity Group at Syracuse University.
B.S. was supported by a QEII Fellowship from the Australian Government.

\bigskip
\noindent
{\bf 7. References}
\medskip
\frenchspacing
\item{[1]} D. Kramer, H. Stephani, M. MacCallum and E. Herlt,
Exact Solutions of Einsten's Field Equations, (Cambridge University
Press, 1980).
\item{[2]} E. Verdaguer, Phys. Rep. {\bf 229} (1993).
\item{[3]} R. Geroch, J. Math. Phys. {\bf 12} (1971) 918;
{\bf 13} (1972) 394.
\item{[4]} C. M. Cosgrove, J. Math. Phys. {\bf 21} (1980) 2417;
{\bf 22} (1981) 2624; {\bf 23} (1982) 615.
\item{[5]} W. Kinnersley, J. Math. Phys. {\bf 18} (1977) 1529;
\item{   } W. Kinnersley and D. M. Chitre;
J. Math. Phys. {\bf 18} (1977) 1538; {\bf 19} (1978) 1926;
{\bf 19} (1978) 2037.
\item{[6]} J. Hauser and F. J. Ernst, Phys. Rev. D {\bf 20} (1979) 362;
J. Math. Phys. {\bf 21} (1980) 1126; {\bf 22} (1981) 1051.
\item{[7]} B. K. Harrison, Phys. Rev. Lett. {\bf 41} (1978) 1197;
Phys. Rev. D {\bf 21} (1980) 1695.
\item{[8]} V. A. Belinski\v i and  V. E. Zakharov, Sov. Phys. JETP
{\bf 48} (1978) 985; \z \z \z \break
{\bf 50} (1979) 1;
\item{   } V. A. Belinski\v i, Sov. Phys. JETP {\bf 50}  (1979) 623;
\item{   } G. A. Alekseev and V. E. Belinski\v i, Sov. Phys. JETP {\bf 51}
(1980) 655.
\item{[9]} D. W. Kitchingham, Class. Quant. Grav. {\bf 1} (1984) 677;
{\bf 3} (1986) 133.
\item{[10]} V. E. Zakharov and A. B. Shabat, Funct. Anal. Appl.
{\bf 8} (1974) 266; {\bf 13} (1979) 166;
\item{    } V. E. Zakharov, The Method of Inverse Scattering Problem,
Lecture Notes in Mathematics, (Springer Verlag, N.Y., 1978).
\item{[11]} L. D. Faddeev, Les Houches, Session XXXIX, 1982,
ed. J.-B. Zuber and R. Stora, (Elsevier Science Publishers B. V., 1984);
\item{   } L. D. Faddeev and L. A. Takhtajan, Hamiltonian Methods in the
Theory of Solitons, (Springer-Verlag, 1987).
\item{[12]} R. H. Gowdy, Ann. Phys. {\bf 83}  (1974) 203,
\item{   } P. T. Chru\' sciel, Ann. Phys. {\bf 202} (1990) 100.
\item{[13]} R. T. Jantzen, Il Nuovo Cimento {\bf 58B} (1980) 287.
\item{[14]} V. A. Belinski\v i and M. Francaviglia, Gen. Rel. Grav.
{\bf 14} (1982) 213.
\item{[15]} N. Manojlovi\' c and G. Stephens, in preparation.
\item{[16]} B. J. Carr and E. Verdaguer, Phys. Rev. D {\bf 28} (1983) 2995;
\item{    } E. Verdaguer, Gen. Rel. Grav. 18 (1986) 1045.
\item{[17]} V. I. Smirnov, A Course of Higher Mathematics Vol. II,
(Pergamon Press, 1964).
\item{[18]} R. Loll, N. Manojlovi\' c and G. Stephens, work in progress.
\item{[19]} R. Loll, N. Manojlovi\' c, G. Mena Marug\' an and G. Stephens,
work in progress.

\bye